\documentclass[british,english]{aa}
\usepackage[T1]{fontenc}
\usepackage{url}
\usepackage{amsmath}
\usepackage{amssymb}
\usepackage{graphicx}

\makeatletter

\usepackage{txfonts}
\bibpunct{(}{)}{;}{a}{}{,} 

\makeatother

\usepackage{babel}
\begin{document}

\title{Flat-relative optimal extraction\thanks{Based on data obtained from the ESO Science Archive Facility under request number ZECHMEISTER73978. Based on observations made with the HARPS instrument on the ESO 3.6 m telescope under programme ID 074.D-0380.}}

\subtitle{A quick and efficient algorithm for stabilised spectrographs}

\author{M.~Zechmeister\inst{1}\and G.~Anglada-Escud\'{e}\inst{1,2}\and
A.~Reiners\inst{1}}

\institute{Institut f\"{u}r Astrophysik, Georg-August-Universit\"{a}t, Friedrich-Hund-Platz
1, 37077 G\"{o}ttingen, Germany\\ \email{zechmeister@astro.physik.uni-goettingen.de}\and
Astronomy Unit, Queen Mary University of London, Mile End Road, London
E1 4NS, UK}

\date{Received / Accepted}

\abstract{Optimal extraction is a key step in processing the raw images of
spectra as registered by two-dimensional detector arrays to a one-dimensional
format. Previously reported algorithms reconstruct models for a mean
one-dimensional spatial profile to assist a properly weighted extraction.}{We
outline a simple optimal extraction algorithm including error propagation,
which is very suitable for stabilised, fibre-fed spectrographs and
does not model the spatial profile shape.}{A high signal-to-noise,
master-flat image serves as reference image and is directly used as
an extraction profile mask. Each extracted spectral value is the scaling
factor relative to the cross-section of the unnormalised master-flat
which contains all information about the spatial profile as well as
pixel-to-pixel variations, fringing, and blaze. The extracted spectrum
is measured relative to the flat spectrum.}{Using echelle spectra
of the HARPS spectrograph we demonstrate a competitive extraction
performance in terms of signal-to-noise and show that extracted spectra
can be used for high precision radial velocity measurement.}{Pre-
or post-flat-fielding of the data is not necessary, since all spectrograph
inefficiencies inherent to the extraction mask are automatically accounted
for. Also the reconstruction of the mean spatial profile by models
is not needed, thereby reducing the number of operations to extract
spectra. Flat-relative optimal extraction is a simple, efficient,
and robust method that can be applied easily to stabilised, fibre-fed
spectrographs.}


\authorrunning{Zechmeister et al.}

\maketitle

\section{Introduction}

Spectra of astronomical objects provide a wealth of information, and
the increasing need for higher precision has led to the development
of very stable and fibre-fed spectrographs. A prospering example is
the search for exoplanets with Doppler spectroscopy. Cross-dispersed,
high-resolution echelle spectrographs are usually employed to measure
radial velocities at a precision of 1 m/s, which corresponds to about
1/1000 of a pixel. Therefore careful calibration, as well as high
mechanical, thermal, and pressure stability, is essential.

If aiming for high precision, not only is the hardware important,
but also the software algorithms to process the images. Typically,
a reduction of echelle spectra consists of several steps: bias subtraction,
dark subtraction, scattered light subtraction, flat-fielding, extraction
to 1D, deblazing, wavelength calibration, order merging, flux normalisation
(e.g. \citealp{Baranne1996}). In this work we focus on extraction,
which is a crucial step in image processing. A widely used method
is the so-called optimal extraction \citep{Horne1986} and its variants
(e.g. \citealp{Marsh1989,Piskunov2002}, Table~1), which basically
scales 1D cross-sectional profiles to the imaged spectrum, and the
scaling factor is the best flux estimate. Additionally, most algorithms
try to model and reconstruct the spatial profile/slit function with
polynomial, Gaussian, or other smooth functions. However, for stabilised
spectrographs, the order profiles and positions are object- and time-independent,
which simplifies the extraction; in particular, there is not necessarily
any need to model the spatial profile with empirical functions. We
exploit this circumstance, and we derive and test our concept of flat-relative
optimal extraction.

\begin{table*}
\caption{\selectlanguage{british}%
\label{tab:OXTmethods}\foreignlanguage{english}{Optimal extraction
algorithms.}\selectlanguage{english}
}

\begin{tabular}{lll}
\hline 
\hline Reference & Cross-section model & Comment\\
\hline 
\citet{Hewett1985} & average along dispersion & assumes no order tilt\\
\citet{Horne1986}; \citet{Robertson1986} & polynomials along dispersion & assumes small order tilt\\
\citet{Urry1988} & Gaussian function & \\
\citet{Marsh1989} & coupled polynomials along dispersion & employs spatial subpixel grid\\
\citet{Piskunov2002} & penalised functions in spatial direction & employs spatial subpixel grid\\
this work & (master flat) & requires stabilised spectrograph\\
\hline 
\end{tabular}
\end{table*}

\section{Principle of flat-relative optimal extraction (FOX)}

In the following, we assume that theimage processing steps (e.g.,
bias, dark, and background subtraction) preceding extraction are properly
done and that, in the image, the main dispersion is oriented in a
more or less horizontal direction ($x$) and the cross-dispersion
(echelle orders) in a vertical direction ($y$).

A spectrograph consists of dispersive elements and a camera that images
the slit or the fibre exit to wavelength-dependent positions and shapes.
The observed light distribution $S(x,y)$ is a convolution of the
input spectrum $s(\lambda)$ with a wavelength-dependent instrumental
point spread function (iPSF) of the spectrograph $\Psi(x,y,\lambda)$.
Therefore a model $\hat{S}(x,y)$ for the observed spectrum can be
formulated as 
\begin{equation}
\hat{S}(x,y)=\Psi(x,y,\lambda)\otimes s(\lambda)\label{eq:2Dconvmodel}
\end{equation}
where wavelength $\lambda$ and the positions $x$ and $y$ in the
detector plane are continuous variables (the hat indicates the model
or best estimate, while without the hat it indicates observations
with noise: $S(x,y)=\hat{S}(x,y)+\sigma(x,y)$). Furthermore, the
spectrum is recorded and binned by detector pixels. Hence it is convenient
to use an effective point spread function (ePSF, \citealp{Anderson2000})
and a discrete version of Eq.~(\ref{eq:2Dconvmodel}) as in \citet{Bolton2010}
\begin{equation}
\hat{S}_{x,y}=\sum_{\lambda}\Psi_{x,y,\lambda}s_{\lambda}\label{eq:2dmodel}
\end{equation}
where $x$ and $y$ now correspond to pixel indices, and a finite
integration over wavelength $\lambda$ still has to be done. The calibration
matrix $\Psi_{x,y,\lambda}$ tabulates the effective response function
of the spectrograph and detector. In \citet{Bolton2010}, Eq.~(\ref{eq:2dmodel})
serves as model for ``perfect'' extraction.

Optimal extraction uses the column numbers as extraction grid ($\lambda\rightarrow x$)
and basically assumes 1D slit functions; i.e., any input wavelength
$\lambda$ that corresponds to the pixel $x$ is imaged only to that
column, meaning only pixels in the spatial direction $y$ are affected,
but no neighbouring columns. Under these circumstances the calibration
matrix can be separated as
\begin{equation}
\Psi_{x,y,\lambda}=\delta_{x,\lambda}\psi_{\lambda,y}\label{eq:OXTassumption}
\end{equation}
 where $\mbox{\ensuremath{\delta}}_{x,\lambda}$ is the Kronecker
delta%
\footnote{$\delta_{ij}=\begin{cases}
1 & \mathrm{if}\, i=j\\
0 & \mathrm{else}
\end{cases}$%
} and $\psi_{\lambda,y}$ the wavelength-dependent cross-section. Then
the model image in Eq.~(\ref{eq:2dmodel}) simplifies to
\begin{equation}
\hat{S}_{x,y}=\psi_{x,y}s_{x}\,.\label{eq:FOXmodel}
\end{equation}
The response $\psi_{x,y}$ could be measured directly with a uniform
input spectrum $s_{x}=1$, but it is much more challenging to determine
$\Psi_{x,y,\lambda}$. In practice, exposures of flat lamps $F_{x,y}$
are usually taken as part of regular calibration sets. Those flat
exposures have high signal-to-noise ratios (S/Ns), and it can be further
increased in master-flats by coadding many flat exposures. This means
that the errors are negligible compared to science exposures and that
we can set $\hat{F}_{x,y}\simeq F_{x,y}$. The spectrum of a flat
lamp $f_{x}$ is generally not uniform overall, but continuous and
featureless, and it varies slowly varying with wavelength. If $f_{x}$
is known, we measure the response as $\psi_{x,y}=\frac{F_{x,y}}{f_{x}}$
directly. However, $f_{x}$ is usually not known in advance and actually
it cannot be measured from the flat exposure alone. It should be characterised
externally, either in advance by another flux-calibrated instrument
or afterwards by observations of standard stars (with known spectra)
with the same instrument as part of the flux calibration step. Since
both might not be available, we extract the science spectrum $s_{x}$
relative to the flat spectrum $f_{x}$ and write for the model 
\begin{equation}
\hat{S}_{x,y}=F_{x,y}\frac{s_{x}}{f_{x}}\,.\label{eq:FOXmodel_used}
\end{equation}

\begin{figure}
\centering

\includegraphics[bb=36bp 296bp 519bp 495bp,clip,width=1\linewidth]{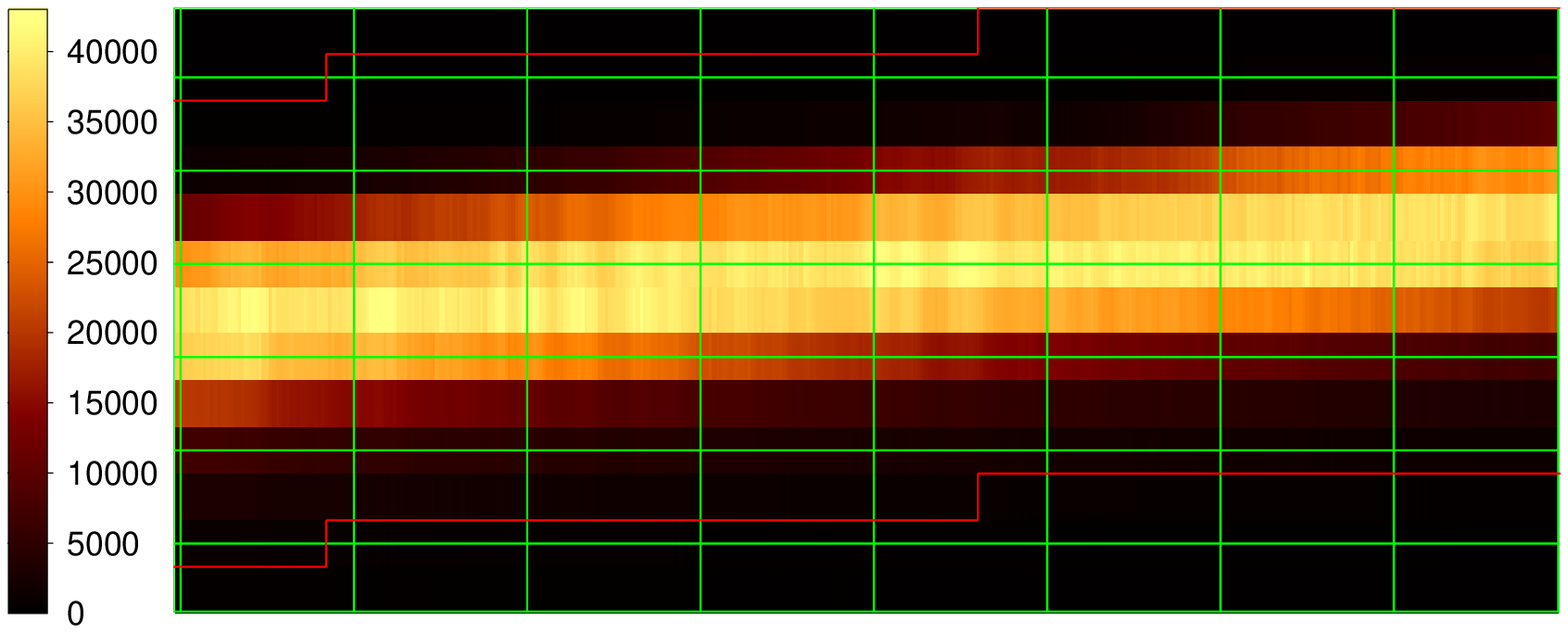}

\includegraphics[bb=36bp 296bp 519bp 495bp,clip,width=1\linewidth]{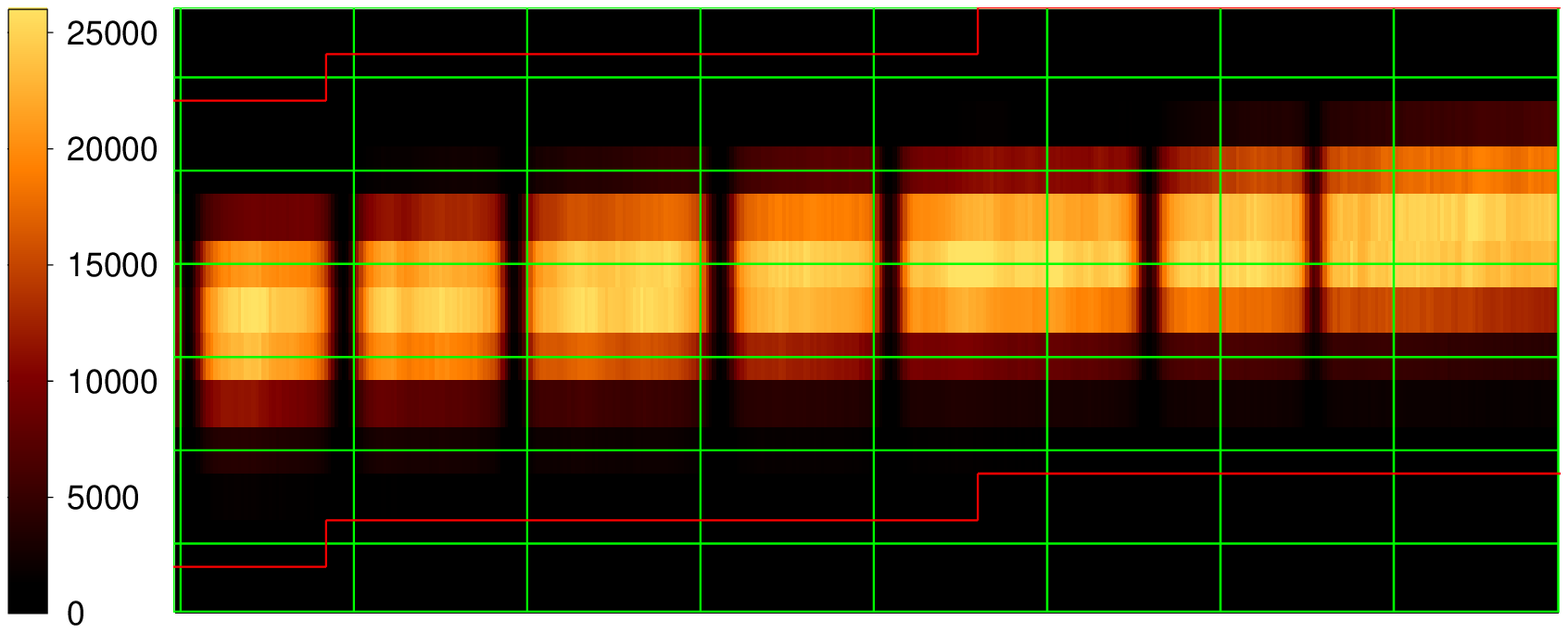}

\includegraphics[bb=50bp 50bp 673bp 390bp,clip,width=1\linewidth]{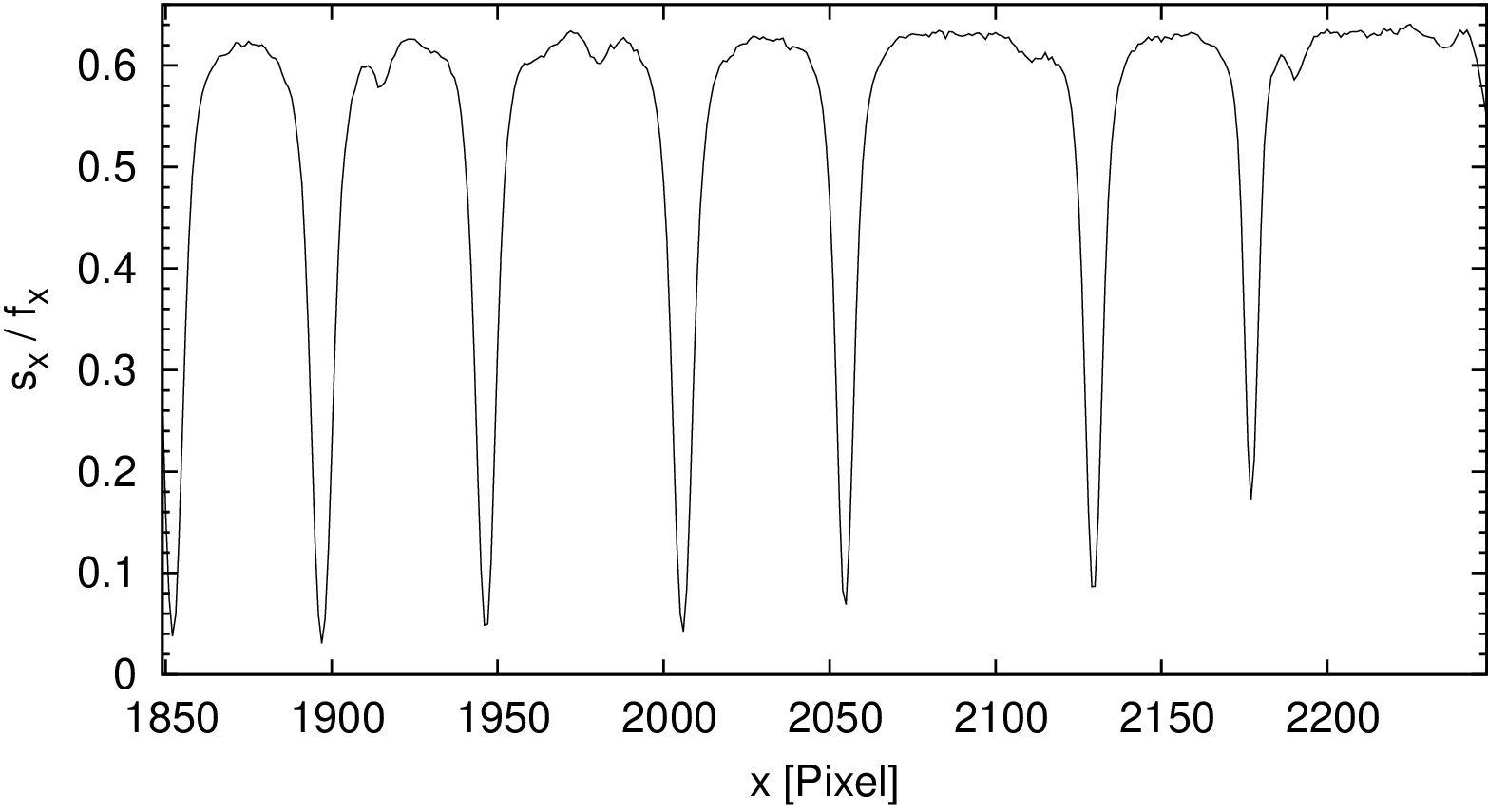}

\caption{\label{fig:FOX-principle}Principle of flat-relative optimal extraction
(FOX). Each of the two upper panels shows a 400\,pix\,$\times$\,13\,pix
section of the reddest HARPS order (fibre A). Pixel-to-pixel variations,
fringing, order tilt/curvature, and the spatial profile are present
in both the master-flat $F_{x,y}$ (\emph{top}) and the stellar raw
spectrum $S_{x,y}$ which has additional (telluric) absorption lines
(\emph{middle}). The extracted spectrum $s_{x}/f_{x}$ (\emph{bottom})
is the scaling factor between both spectra for each column within
the extraction mask (red lines).}
\end{figure}

We obtain the spectrum $\frac{s_{x}}{f_{x}}$ by minimising the residuals
bet\-ween observations $S_{x,y}$ and model $\hat{S}_{x,y}$, i.e.
solving the linear least-square problem 
\begin{equation}
\chi^{2}\approx\sum_{x,y}w_{x,y}\left[S_{x,y}-F_{x,y}\frac{s_{x}}{f_{x}}\right]^{2}=\mathrm{minimum}\label{eq:FOXchisqr}
\end{equation}
where $w_{x,y}$ are the weights for each pixel using a noise model~$\sigma_{x,y}$
(e.g. photon and readout noise; see Sect.~\ref{sec:Noise-model}).
These weights may also include a map $M_{x,y}$ masking bad pixels
and pixels outside the extraction aperture ($w_{x,y}=\frac{M_{x,y}}{\sigma_{x,y}^{2}}$).
Setting the derivative of $\chi^{2}$ with respect to $\frac{s_{x}}{f_{x}}$
equal to zero, we get a set of decoupled equations with the solution
at each position

\begin{equation}
r_{x}\equiv\frac{s_{x}}{f_{x}}=\frac{\sum_{y}w_{x,y}F_{x,y}S_{x,y}}{\sum_{y}w_{x,y}F_{x,y}F_{x,y}}\,,\label{eq:FOX}
\end{equation}
where $\frac{s_{x}}{_{fx}}$ is the best fitting amplitude at each
spectral bin $x$. Figure~\ref{fig:FOX-principle} illustrates the
principle of flat-relative optimal extraction (FOX).

Equation~(\ref{eq:FOX}) is quite similar to the well known optimal
extraction equation (e.g. \citealp{Horne1986}). The difference is
that we extract the spectrum $s_{x}$ relative to the flat spectrum
$f_{x}$, and $F_{x,y}$ is not normalised and includes the natural
spatial profile and all flat-field effects.

\section{\label{sec:Noise-model}Noise model and error estimation}

The pixels on the detector (and therefore the extracted values $s_{x}/f_{x}$)
are affected by noise $\sigma_{x,y}$. This consists mainly of photon
noise $\sigma_{\mathrm{ph}}^{2}=\hat{S}_{_{x,y}}$ (if both $\sigma_{\mathrm{ph}}$
and $\hat{S}_{x,y}$ are in units of photon counts)%
\footnote{If a sky- or stray light background was subtracted from the image
before, its contribution to photon noise should be also taken into
accounted.%
} and detector noise (e.g. the readout noise $\sigma_{\mathrm{rn}}^{2}$
of a CCD). Therefore, a simple noise model for each pixel is, for
instance (now all units in ADU), 
\begin{equation}
\sigma_{x,y}^{2}=\sigma_{\mathrm{rn}}^{2}+g\hat{S}_{_{x,y}}\label{eq:noisemodel}
\end{equation}
where constant gain $g$ (which converts the number of photoelectrons
to ADU)%
\footnote{There are different definitions of the gain in the literature. Here
we use $S[\mathrm{ADU]}=g[\mathrm{ADU}/e^{-}]\cdot S[e^{-}]$. The
photon noise in units of ADU is $\sigma_{\mathrm{ph}}[\mathrm{ADU]}=g[\mathrm{ADU}/e^{-}]\cdot\sigma_{\mathrm{ph}}[e^{-}]=g[\mathrm{ADU}/e^{-}]\cdot\sqrt{S[e^{-}]}=\sqrt{g[\mathrm{ADU}/e^{-}]\cdot S[\mathrm{ADU}]}$.%
} and constant readout noise $\sigma_{\mathrm{rn}}^{2}$ are assumed.

With a given noise model, one can predict the extraction uncertainties
through error propagation of Eq.~(\ref{eq:FOX})
\begin{align}
\mathrm{var}(r_{x})=\epsilon_{r_{x}}^{2} & =\sum_{y}\left(\frac{\partial r_{x}}{\partial S_{x,y}}\right)^{2}\sigma_{x,y}^{2}=\sum_{y}\left(\frac{F_{x,y}/\sigma_{x,y}^{2}}{\sum_{y'}w_{x,y'}F_{x,y'}^{2}}\right)^{2}\sigma_{x,y}^{2}\nonumber \\
 & =\frac{1}{\left(\sum_{y}F_{x,y}^{2}/\sigma_{x,y}^{2}\right)^{2}}\sum_{y}F_{x,y}^{2}/\sigma_{x,y}^{2}=\frac{1}{\sum_{y}F_{x,y}^{2}/\sigma_{x,y}^{2}}\label{eq:error_abs}
\end{align}
 (similar to \citealp{Horne1986}). We also see that the relative
error,
\begin{equation}
\frac{\epsilon_{r_{x}}}{r_{x}}=\frac{\sqrt{\sum_{y}w_{x,y}F_{x,y}^{2}}}{\sum_{y}w_{x,y}F_{x,y}S_{x,y}}\,,\label{eq:error_rel}
\end{equation}
 is smaller in regions with high flux $S_{x,y}$ and is scaling invariant
with respect to the flat image $F_{x,y}$, because multiplying $F_{x,y}$
by a constant (as happens for a longer flat exposure time) will cancel
out in the ratio in Eq.~(\ref{eq:error_rel}).

The a priori predicted uncertainty in Eq.~(\ref{eq:error_abs}) does
not account for the goodness of fit and thus does not indicate profile
or noise model mismatches. For this reason, the error estimates from
the $\chi^{2}$ modelling are often rescaled by $\chi_{\mathrm{red}}^{2}$(the
reduced $\chi^{2}$) to provide a posteriori estimated errors
\begin{equation}
\sigma_{s_{x}/f_{x}}=\chi_{\mathrm{red}}\cdot\epsilon_{s_{x}/f_{x}}\label{eq:error-post}
\end{equation}
where $\chi_{\mathrm{red}}=\sqrt{\frac{1}{N-\nu}\,\chi^{2}}$ with
$\chi^{2}$ from the global fit in Eq.~(\ref{eq:FOXchisqr}). The
number of degrees of freedom $N-\nu$ in the denominator is given
by the number of unmasked pixels $N=\sum_{x,y}M_{x,y}$ within the
extraction aperture and the number of fitted parameters $\nu$ (extracted
spectral values).

One can also consider another, second choice for the scaling factor,
namely $\chi_{\mathrm{red},x}$ to be taken from the individual cross-section
fits, i.e. $\chi_{\mathrm{red},x}^{2}=\frac{1}{N_{x}-1}\,\chi_{x}^{2}$
where only one free parameter is left ($\nu=1$), $\chi_{x}^{2}$
is the weighted sum of the residuals only in column $x$ and $N_{x}=\sum_{y}M_{x,y}$
is the number of unmasked pixels in that column. However, for fibre-fed
spectrographs, the extraction aperture is only a few pixels ($N_{x}\sim10$)
wide. This gives low number statistics making $\chi_{\mathrm{red},x}$
itself very uncertain (in contrast to $\chi_{\mathrm{red}}^{2}$ which
is approximately the average of all $\chi_{\mathrm{red},x}^{2}$)%
\footnote{$\left\langle \chi_{\mathrm{red},x}^{2}\right\rangle =\frac{1}{\nu}\sum_{x}\frac{1}{N_{x}-1}\,\chi_{x}^{2}\approx\frac{1}{\nu}\frac{1}{\left\langle N_{x}\right\rangle -1}\sum_{x}\chi_{x}^{2}=\frac{1}{\nu\left\langle N_{x}\right\rangle -\nu}\,\chi^{2}=\chi_{\mathrm{red}}^{2}$.
The approximation becomes an equality if the aperture width is the
same for all columns, and no pixels are rejected ($N_{x}=\mathrm{const}$,
$N\approx\nu\left\langle N_{x}\right\rangle $).%
}. In particular, a considerable number of $\chi_{\mathrm{red},x}$
will occur with values much less than one, when fitting the profiles
along the dispersion axis. However, the predicted errors should not
be smaller than the fundamental limit (e.g. photon noise).

As discussed in \citet{Horne1986}, cosmic ray hits can be efficiently
detected and removed with optimal extraction. Those cosmics distort
the profile and can be identified as significant outliers by means
of the noise model, e.g. by setting an upper threshold
\begin{equation}
S_{x,y}-\hat{S}_{x,y}>\kappa\sqrt{\mathrm{var}(S_{x,y}-\hat{S}_{x,y})}=\kappa\cdot(\sigma_{x,y}^{2}-F_{x,y}^{2}\epsilon_{s_{x}/f_{x}}^{2})\,,\label{eq:kapsigclip}
\end{equation}
(see Appendix~\ref{sec:Appendix} for a derivation of this equation),
where the clipping value is typically $\kappa\sim3-5$ to reject cosmics.
A lower threshold could be set as well, e.g. to detected unmasked
cold pixels. Moreover, other or additional criteria are also used
\citep{Baranne1996}. The outliers are masked and the extraction process
is repeated.

We note that the variance in the residual image $S_{x,y}-\hat{S}_{x,y}$
is generally not just the pixel variance $\sigma_{x,y}^{2}$. In contrast
to \citet{Horne1986} the correction term $F_{x,y}^{2}\epsilon_{s_{x}/f_{x}}^{2}$
should be applied in Eq.~(\ref{eq:kapsigclip}). Figuratively, the
observed dispersion in the residuals will be noticeably smaller than
$\sigma_{x,y}^{2}$, because the number of pixels within the aperture
is, as already mentioned, small and the profile is usually not uniform,
but concentrates most flux in even fewer pixels (\textasciitilde{}3-4~pixels
in fibre-fed spectrographs). Imagine a concentrated profile, where
one pixel has a very high weight. This pixel dominates the fit, thus
forcing its own residual to zero. As an example, in Fig.~\ref{fig:FOX-principle}
the pixel in the profile centre contains about 25\% of the total profile
flux, and the observed dispersion in this pixel will be smaller than
$\sigma_{x,y}$ by a factor of 0.56 in case of readout dominated noise
(Eq.~(\ref{eq:res_var_readout})) and by 0.75 in case of pure photon
noise (Eq.~(\ref{eq:res_var_photon})).

\begin{table*}
\caption{Efficiency components of typical echelle spectrographs visible in
flat-fields.}

\begin{tabular}{lccl}
\hline 
\hline  & Amplitude scale & Size scale & Source\\
\hline 
$\varepsilon_{\mathrm{pixel}}$ & 1\,--\,4\% & 1\,pix & pixel efficiency, pixel size, and quantum efficiency (CCD)\\
$\mathrm{\varepsilon_{\mathrm{fringe}}}$ & 0\,--\,20\% & 20\,pix & fringing, interference pattern\\
$\mathrm{\varepsilon_{\mathrm{blaze}}}$ & 0\,--\,100\% & 500\,pix & blaze function (echelle grating)\\
$\mathrm{\varepsilon_{\mathrm{\lambda}}}$ & 0\,--\,100\% & orders & wavelength-dependent efficiency of spectrograph and detector\\
\hline 
\end{tabular}
\end{table*}

\section{Conceptual comparison with other optimal extraction implementations}

Numerous optimal extraction algorithms (hereafter OXT) exists that
are listed in Table~\ref{tab:OXTmethods}. Usually, they have to
assume a slowly varying spatial profile and differ in the reconstruction
method for this profile. Using the spatial profiles of many columns
a mean high S/N model is created. Since OXT aims to estimate ``absolute''
counts%
\footnote{Therefore the blaze function and wavelength-dependent efficiencies
are not corrected at this stage, and usually $s_{x}\cdot\varepsilon_{\mathrm{blaze},x}\cdot\varepsilon_{\lambda,x}$
is extracted (instead of $s_{x}$).%
}, the profile functions $p_{x,y}$ are normalised to unity ($\sum_{y}p_{x,y}=1$).

If there is no or slow order tilt, the profile might be reconstructed
by averaging \citep{Hewett1985} or fitting low-order polynomials
in the dispersion direction \citep{Horne1986}, respectively. For
echelle spectrographs, this situation occurs only in small sections
of the orders, but not in general. For larger tilts, the order trace
(which describes the vertical position of the profile along the $x$
direction) appears more or less explicitly in the profile modelling,
and sub-pixel grids are introduced. Then neighbouring polynomials
in dispersion direction are coupled \citep{Marsh1989}, or recentred
cross-sections are fitted by smoothing (spline-like) functions \citep{Piskunov2002,Bolton2007}.
Those algorithms need a good description of the trace to properly
position the profile model. In particular, for order regions grazing
the detector borders, the order tracing and the application of OXT
algorithms is difficult.

An advantage of OXT is that the profile reconstruction can be done
on the object spectrum itself. This is crucial for slit spectrographs
where the shape and position of the object on the slit and/or detector
can be different in the next exposure or change even during an exposure,
depending on the seeing, guiding, and temperature. The noise in the
mean profiles is then reduced by about the square root of the number
of used columns compared to individual profiles (assuming an absorption
spectrum; however, this approach may have problems with emission line
spectra that have profile information only in a few columns).

The concept of optimal extraction was originally developed for slit
spectrographs, and the application to fibre-fed spectrographs seemed
natural. If the image shape becomes insensitive to seeing and guiding
(as for fibre-fed instruments), one can save the effort of the profile
modelling on the object itself (which can become computationally extensive
as in \citealp{Piskunov2002}) and a less noisy reference profile
model might be taken from a reference image or a flat \citep{Marsh1989,Baranne1996}.
Still, a recentring of the model might be needed to account for sub-pixel
shifts (e.g. a nightly shift of 0.1 pixels was reported by \citeauthor{Baranne1996}
for ELODIE).

If additionally the position of the image is fixed, as for stabilised
instruments without mechanical and thermal flexure, then recentring
also becomes redundant. In this case there is even no longer any need
to model the profile that now can be taken directly from the flat,
therefore FOX does not require any choices for model parameters such
as the polynomial degree, the subpixel grid size, or the amount of
smoothing.

Another benefit of the fixed format is that flat-field effects can
be included in the extraction mask, and in this sense FOX does the
flat-fielding simultaneously. Also from Eq.~(\ref{eq:FOX}), we note
that there is no need for a pixel-wise division by a flat (an operation
one likes to avoid, especially when small numbers occur as in the
wings of the cross-sections).

The classical data reduction with OXT requires an additional pre-
and/or post-flat-fielding, i.e. before and/or after extraction \citep{Baranne1996}.
Flat-fielding has the task of correcting for various multiplicative
efficiency effects that have various causes and scales. Locally, there
are, e.g., pixel-to-pixel variations owing to different sizes and
quantum efficiencies of the detector pixels. Fringe patterns are interference
effects that become serious in the red orders and depend on the properties
of CCD (layer thickness) and the incident light (wavelength, angle,
position). The blaze efficiency along and across the orders depends
on the echelle grating and the cross-disperser. Finally, there is
also a wavelength-dependent efficiency of the detector and optical
elements (e.g. fibre transmission). We may summarise these effects
as $\varepsilon_{x,y}=\varepsilon_{\mathrm{pixel},x,y}\,\varepsilon_{\mathrm{fringe},x,y}\,\varepsilon_{\mathrm{blaze},x}\,\varepsilon_{\lambda,x}$.

Basically, $\varepsilon_{\mathrm{blaze}}$ and $\varepsilon_{\lambda}$
are only functions of $x$ and they could be corrected after extraction.
However, $\varepsilon_{\mathrm{pixel}}$ and $\varepsilon_{\mathrm{fringe}}$
have spatial components and should be corrected before (or, as in
FOX, during) the extraction. If not, the profile mismatches result
in less correct flux estimates, and in particular, residing fringe
patterns lead to strong, rapid, and systematic profile variations
violating the above assumption as needed for the profile modelling.

Data reduction pipelines using OXT create a normalised flat-field
image. Those might be obtained as part of the decomposition of flat
exposure into a normalised flat image, a normalised profile/order
shape, and the blaze function \citep{Piskunov2002}. Another technical
approach is to use, if existing, wider or moving fibres to get a uniform
illumination (similar to a long-slit, \citealp{Baranne1996}), but
fringe patterns are likely not properly captured.

\section{FOX in action}

\begin{figure}
\centering

\includegraphics[clip,width=1\linewidth]{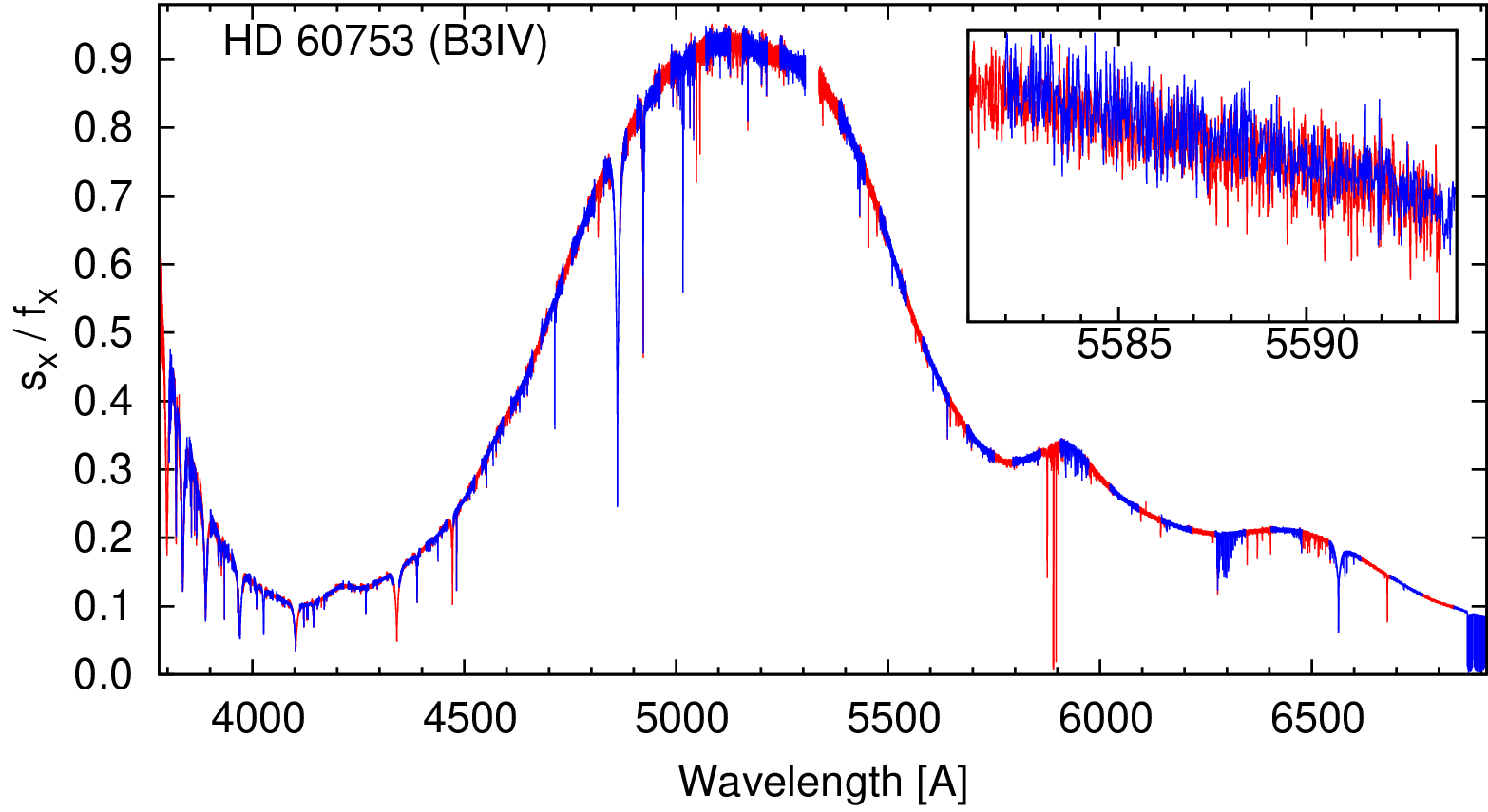}

\includegraphics[clip,width=1\linewidth]{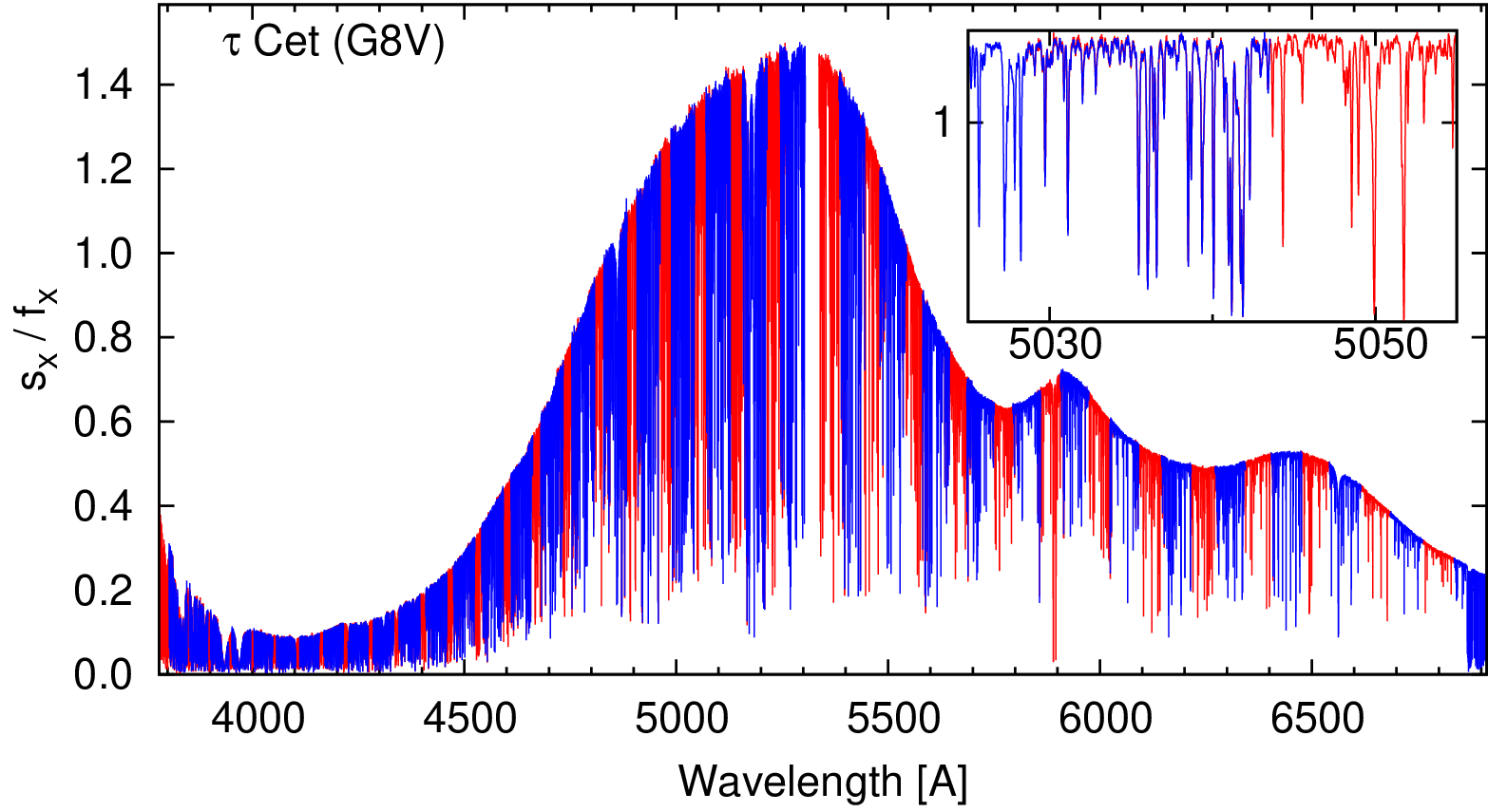}

\caption{\label{fig:Extracted-Echelle-orders}Spectra of HD~60753 and $\tau$~Cet
extracted with FOX. The orders (alternating colours) are not merged.
The spectrum is not flux-corrected, but is relative to the spectrum
of the flat lamp. The inset shows the overlap between two orders.}
\end{figure}

\begin{figure}[!th]
\centering

\includegraphics[clip,width=1\linewidth]{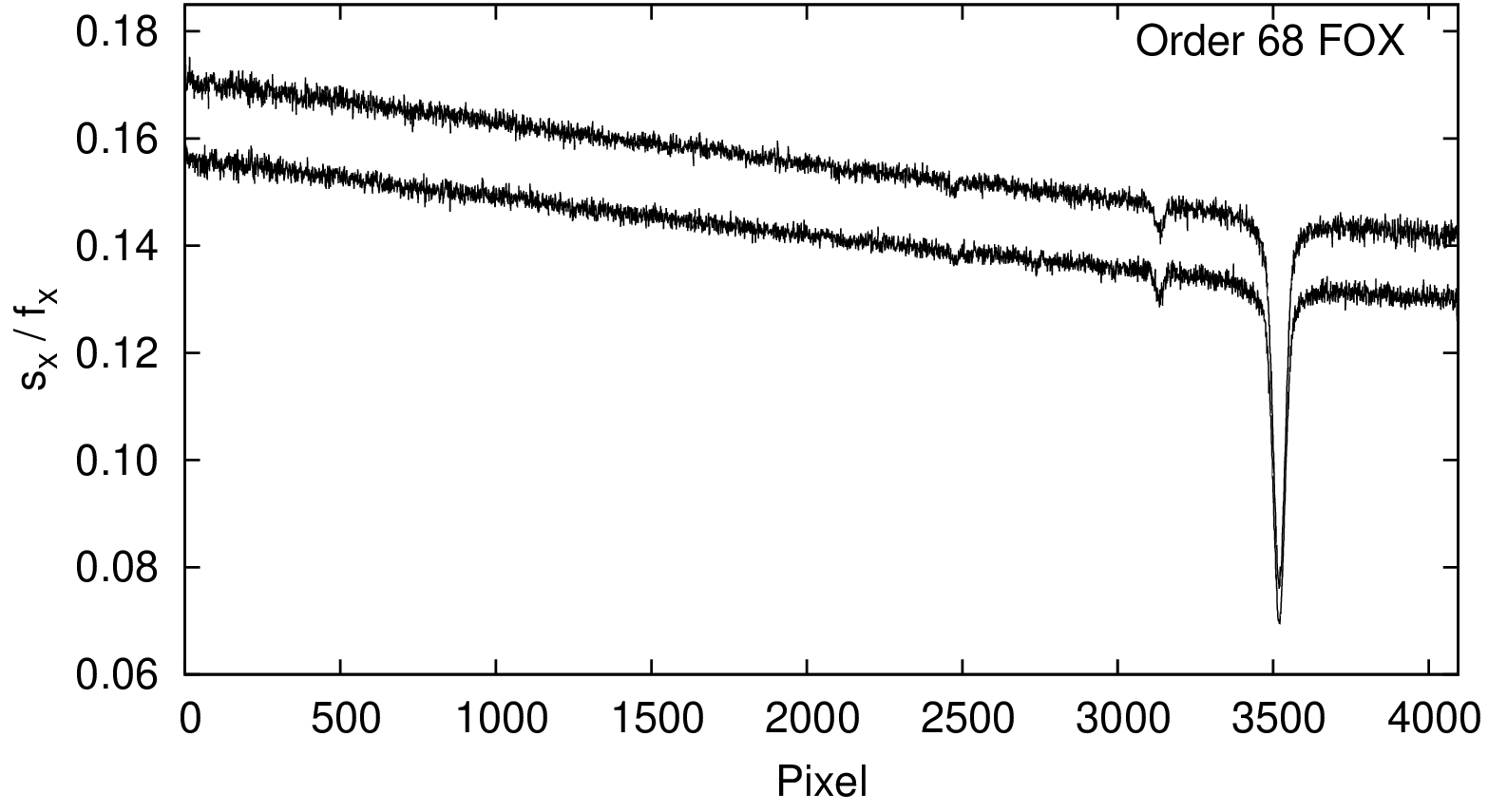}

\includegraphics[clip,width=1\linewidth]{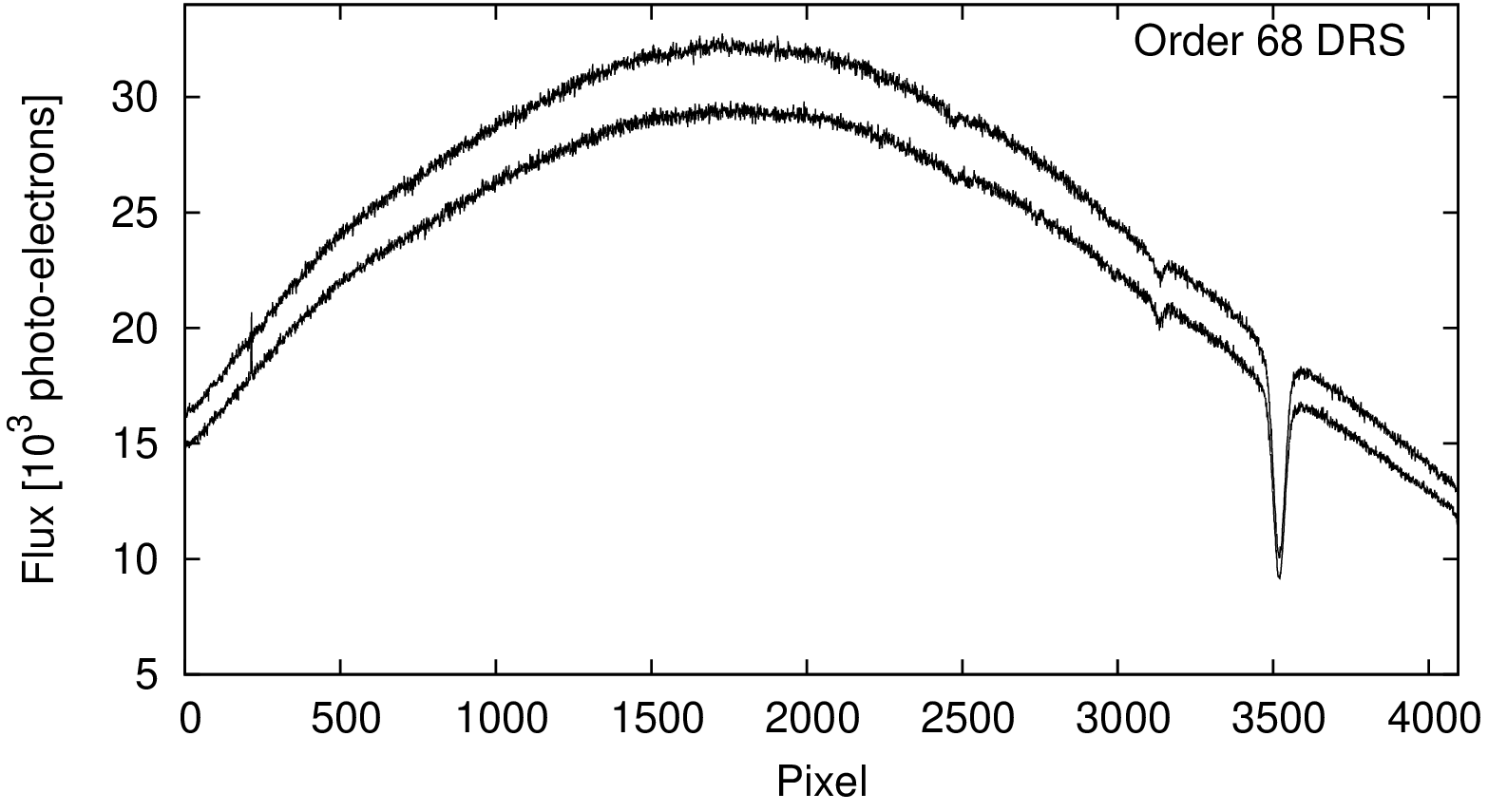}

\includegraphics[bb=50bp 50bp 673bp 390bp,clip,width=1\linewidth]{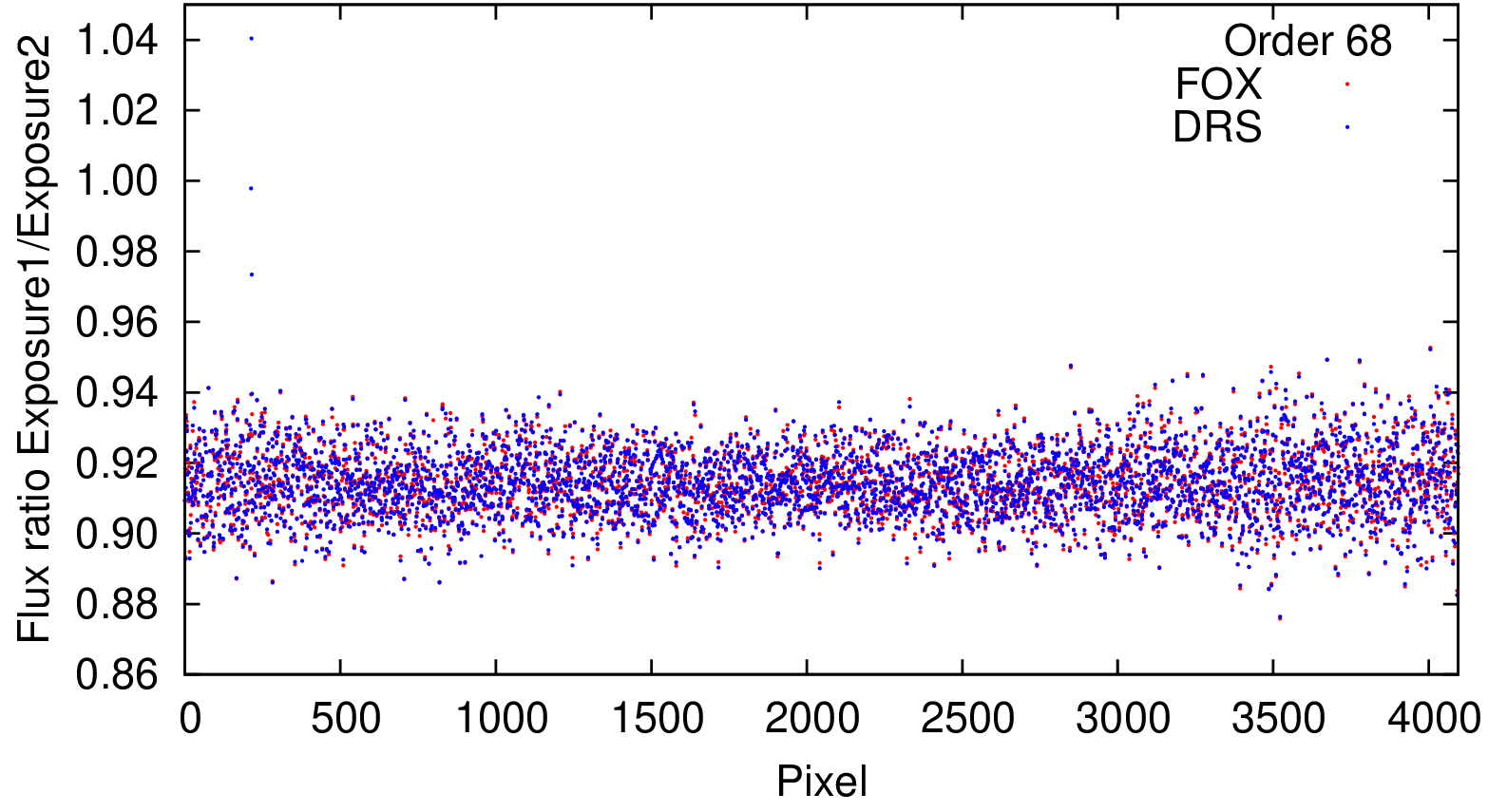}

\includegraphics[bb=50bp 50bp 673bp 390bp,clip,width=1\linewidth]{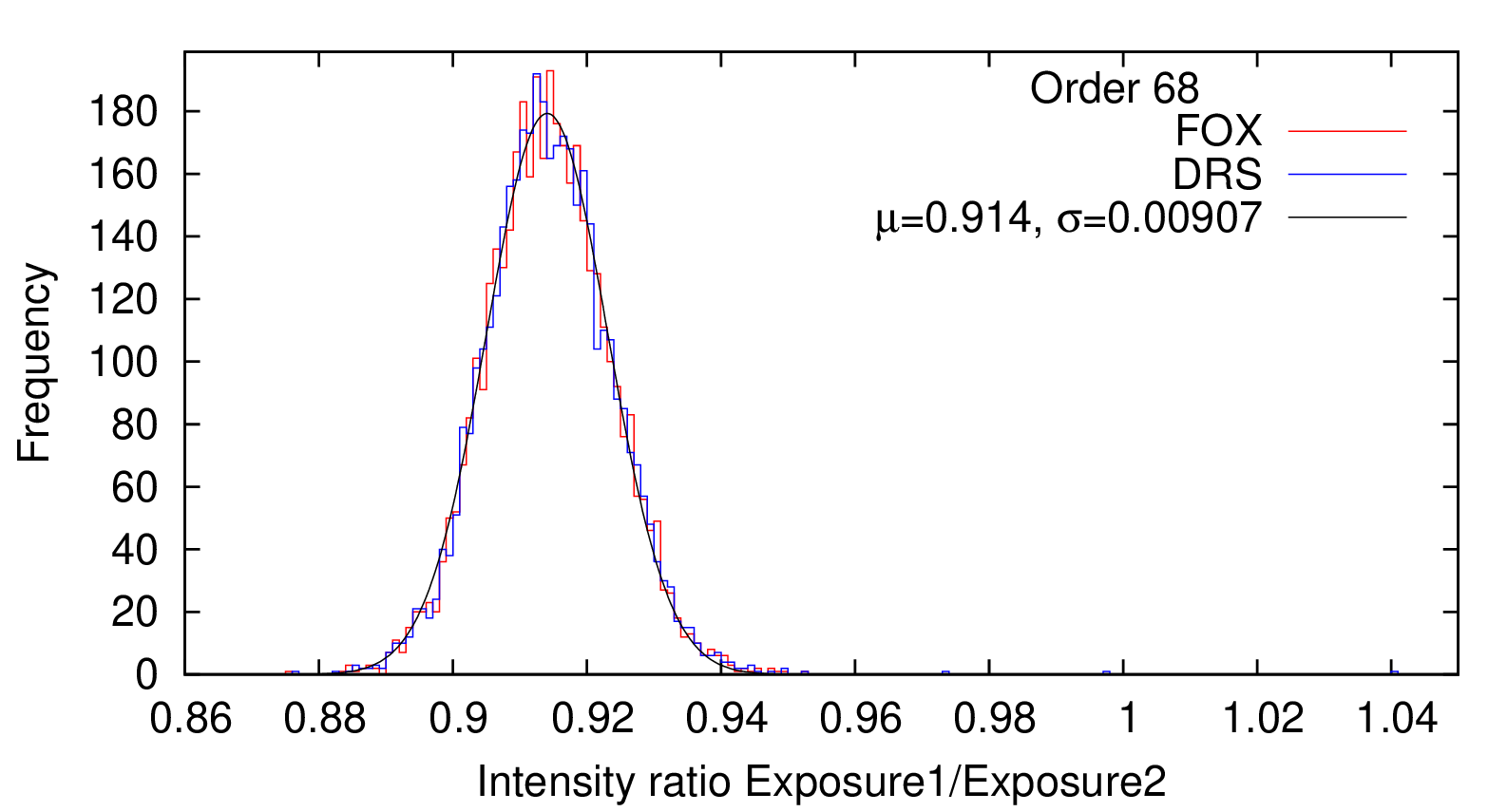}

\caption{\label{fig:FOXvsDRS}Comparison of the extraction quality for two
HARPS spectra of the B3IV star HD~60753 taken in the same night (S/N\textasciitilde{}170).
\emph{Top}: Extraction with FOX. \emph{Second panel}: Spectrum extracted
with the HARPS DRS pipeline (not blaze corrected). \emph{Third panel:}
Ratio of both exposures for FOX (red dots) and DRS (blue dots). \emph{Bottom}:
Histogram of the ratio values. The black line is a Gaussian fit to
the FOX histogram having a centre at 0.914 and a width of 0.00905.}
\end{figure}

To demonstrate the performance of FOX, we extract spectra taken with
HARPS, an echelle spectrograph at the 3.6\,m ESO telescope in La
Silla (Chile), located in a pressure- and temperature-stabilised vacuum
tank \citep{Mayor2003}. It is fed by two fibres from the Cassegrain
focus, a science fibre (A) and a calibration fibre (B) to simultaneously
monitor wavelength drift or sky background. HARPS has a resolving
power of $R=110\,000$ and covers the wavelength range 3800\,--\,6900\,\AA{}
with 72 orders (fibre A) on a 4\,k\,$\times$\,4\,k CCD. Figure~\ref{fig:FOX-principle}
shows a section of a HARPS raw spectrum.

We also choose HARPS observations, because there is an elaborated
pipeline for HARPS called data reduction software (DRS, version 3.5)%
\footnote{\url{http://www.eso.org/sci/facilities/lasilla/instruments/harps/doc/index.html}%
} that already has demonstrated high performance and allows comparison
to our extraction results. The DRS employs the \citet{Horne1986}
algorithm and since version 3.5 uses coadded flat-fields to define
the extraction profile (C. Lovis, Observatoire de Gen\`{e}ve, priv.
comm.). It is here supposed to represent optimal extraction. Both
the raw and reduced spectra are publicly available%
\footnote{\url{http://archive.eso.org/eso/eso_archive_main.html} and \url{http://archive.eso.org/wdb/wdb/eso/repro/form}%
}.

We extracted spectra for a standard star (HD~60753, B3IV) and a solar-like
star ($\tau$~Cet, G8V). We used the REDUCE package of \citet{Piskunov2002}
for the preprocessing: The bias frames were averaged to a master bias;
five flat exposures were bias-subtracted, averaged to a master-flat
and the scattered light was subtracted; one order-location frame (a
flat lamp illuminating only fibre A) was used to define the order
traces of fibre A; from the science images, the master bias and the
scattered light were subtracted (no pre-flat-fielding was performed).
Then we extracted the science spectra with our FOX algorithm with
a spatial extraction width of ten pixels (five whole pixels below
and five above the order trace), $\sigma_{\mathrm{rn}}=5\,\mathrm{ADU}$,
and $g=0.70\,\mathrm{ADU}/\mathrm{e}_{\mathrm{ph}}^{-}$. The wavelength
solutions were taken from the DRS pipeline.

Figure~\ref{fig:Extracted-Echelle-orders} shows the result of the
FOX extraction for two observations. As can be seen, the extracted
orders match in the overlapping regions and seem to be ready to be
merged directly. Order merging, however, is not the topic of this
work and is not necessary for our RV computations below. Merging also
needs some care, since the different resolution and sampling will
lead to mismatches in the absorption and emission features, and in
this way some information is also lost.

\subsection{S/N measurement for the standard star}

To measure the extraction quality, we use two observations of the
standard star HD~60753, taken three hours apart in the night 2007-03-26;
exposures times 9.5\,min, and various methods of estimating S/Ns.
Figure~\ref{fig:FOXvsDRS} displays the extracted aperture number
68 (6613\,--\,6687\AA{}) of the standard star (the prominent absorption
line is HeI).

Using the uncertainties derived from the noise model (Eq.~(\ref{eq:error_abs})),
we estimate for FOX a quadratic mean signal-to-noise of $\mathrm{S/N}=\sqrt{\frac{1}{n}\sum_{x}\frac{r_{x}^{2}}{\epsilon_{r_{x}}^{2}}}=169.6$
(exposure 2: 176.9) per extracted pixel around the central $n=100$
pixels. Since no uncertainties are provided for DRS spectra, we assume
pure photon noise ($\epsilon_{x}=\sqrt{s_{x}}$). Then the count level
in the DRS spectrum implies a mean photon S/N of $\sqrt{\left\langle s_{x}\right\rangle }=170.4$
(exposure 2: 178.6)%
\footnote{Accounting for readout noise from, say, 5 spatial pixels decreases
the S/N values by about $1-\sqrt{(s_{x}+5\sigma_{\mathrm{rn}}^{2}/g)/s_{x}}\sim0.3\%$.%
}. These S/N values are similar and provide fundamental limits. In
this respect, we note that we have $\chi_{\mathrm{red}}=1.08$ for
this order.

Another way to measure the S/N independently of a priori error estimates
is to analyse the scatter in a continuum region. (This is the reason
for choosing a standard star for the comparison.) For the same central
region, the S/N per extracted pixel derived as the ratio of the intensity
mean and the standard deviation ($\mathrm{S/N}=\frac{\left\langle s_{x}\right\rangle }{\left\langle (s_{x}-\left\langle s_{x}\right\rangle )^{2}\right\rangle }$)
is 185 (exposure 2: 163) for FOX and 177 (exposure 2: 141) for DRS.
These numbers already indicate that the extraction qualities are similar.

In a further comparison (also independent of a priori error estimates),
we take the ratio of the two spectra of the standard star and analyse
the scatter. Taking the ratio of both spectra cancels out the remaining
flat spectrum in the case of FOX and the blaze function in the case
of DRS, therefore allowing for a more direct comparison over a wider
range and even pixel-wise. We see in the third panel of Fig.~\ref{fig:FOXvsDRS}
that the ratio values of FOX ($q_{x}=r_{x,\mathrm{exp1}}/r_{x,\mathrm{exp2}}$)
and DRS ($q_{x}=s_{x,\mathrm{exp1}}/s_{x,\mathrm{exp2}}$) have a
similar mean (\textasciitilde{}0.914, constant over the full order)
and are correlated. For both extractions, the scatter appears similar
and barely distinguishable by eye. In both cases the scatter increases
towards the order edges (since the flux decreases due to the blaze,
see second panel of Fig.~\ref{fig:FOXvsDRS}). As before, we measured
a S/N from the mean and the standard deviation of the ratio values
for the central 100 pixels and find a slightly higher S/N for FOX
($\mathrm{S/N}_{q,\mathrm{FOX}}=133$ and $\mathrm{S/N}_{q,\mathrm{DRS}}=129$).

In contrast to the previous method, we can extend this ratio method
over the full order and also use regions with stellar lines (assuming
that the stellar lines are static, i.e. do not vary in shape and position).
A slight, relative shift between both spectra is present owing to
the difference in their barycentric radial velocities (212\,m/s,
$\sim0.25\,\mathrm{pix}$). The slightly increased scatter noticeable
at the position of the strong stellar line is due to this shift as
well as to the lower flux level (lower S/N).

\begin{figure}
\centering

\includegraphics[bb=50bp 50bp 673bp 390bp,clip,width=1\linewidth]{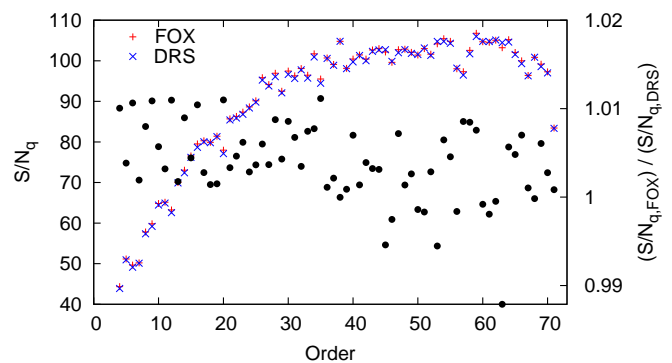}

\caption{\label{fig:snr-FOXvsDRS}Signal-to-noise $\mathrm{S/N}_{q}$ measured
in the spectrum ratio $q_{x}$ of two standard star exposures for
several orders from the extraction FOX (red plus) and DRS (blue crosses).
The quotient of the S/N values between FOX and DRS (black filled circles)
refer to the right axis and are close to unity.}
\end{figure}

The last panel of Fig.~\ref{fig:FOXvsDRS} shows a histogram for
the 4096 ratio values. We see that the ratio values have nearly a
Gaussian distribution, and we now measure the mean and dispersion
more robustly from a Gaussian fit to the histograms. Using $\mathrm{S/N}_{q}=\frac{\mu}{\sigma}$,
we find values of 100.77 (FOX) and 100.87 (DRS) for order 68. The
same procedure was applied to the other orders and the results plotted
in Fig.~\ref{fig:snr-FOXvsDRS}. The overall course of the $\mathrm{S/N}_{q}$
values mostly reflects the instrument efficiency (times the stellar
energy distribution) along the order. Both FOX and DRS extraction
deliver similar $\mathrm{S/N}_{q}$ with quotients close to unity
and deviation of $\lesssim$1\%, whereas in this example FOX provided
slightly higher S/N in the blue orders.

\subsection{RV performance for a Sun-like star}

As a further, less direct, but probably more relevant proxy for the
extraction quality we use radial velocity (RV) measurements. The RV
precision depends on the S/N in the observation and the RV information
content of the stellar spectral lines (the number and amount of gradients;
\citealp{Bouchy2001}). In particular this means that we are probing
the extraction quality in the flanks of spectral lines (rather than
in continuum regions as before).

\begin{figure}
\centering

\includegraphics[width=1\linewidth]{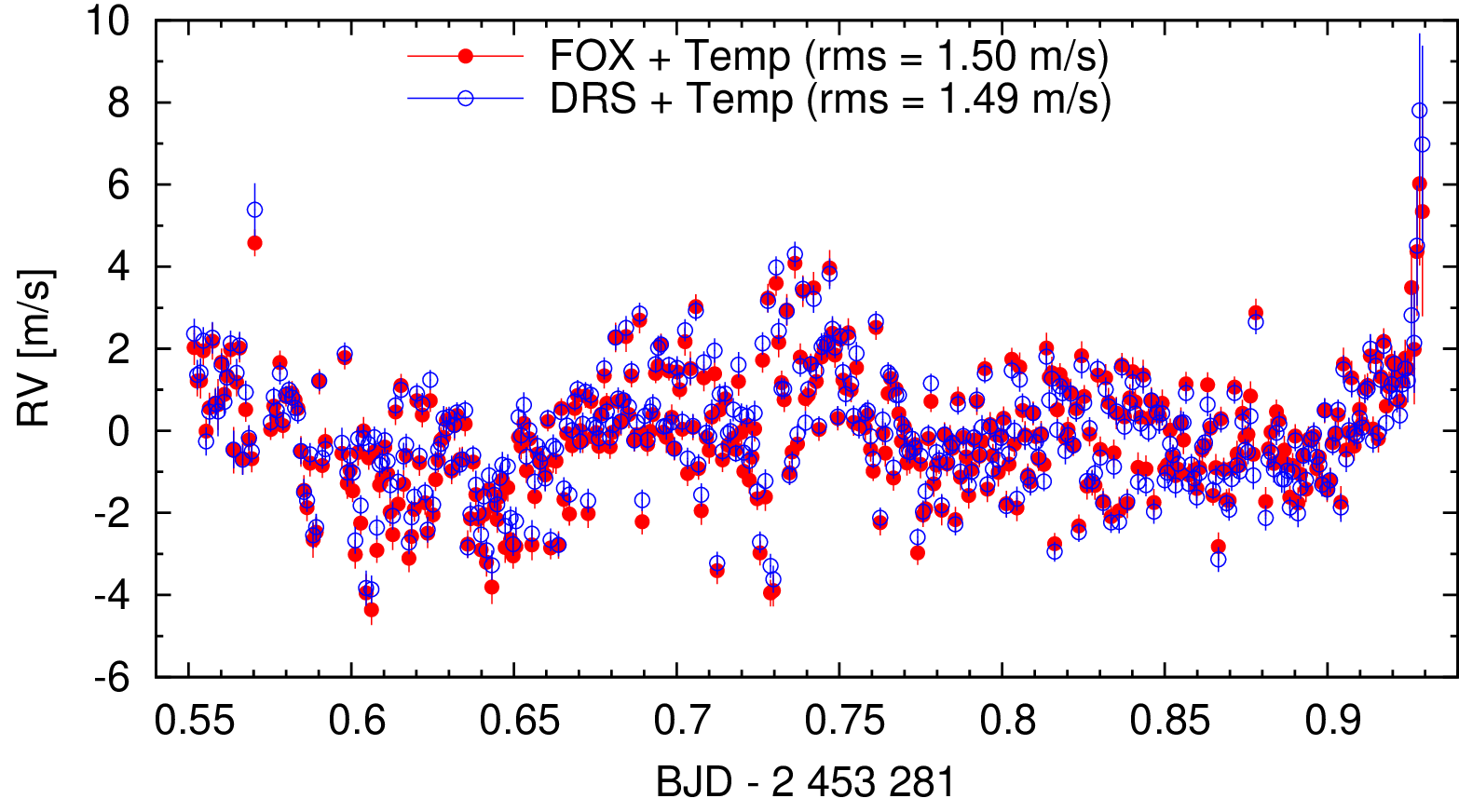}

\includegraphics[width=1\linewidth]{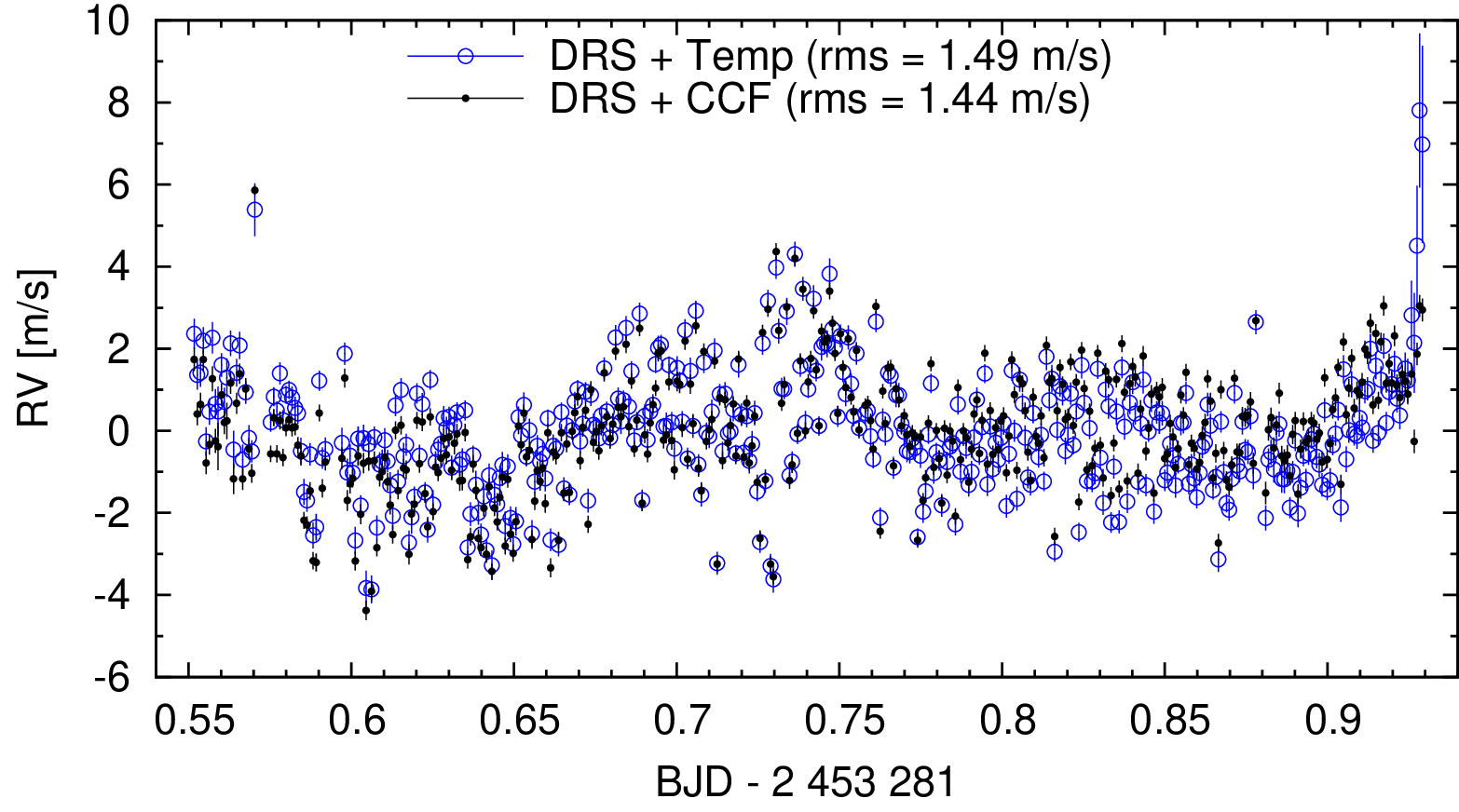}

\caption{\label{fig:RV-TauCet}Radial velocities of $\tau$~Cet in the night
of 2004-10-02 computed with the least square method for FOX and DRS
extraction (\emph{top}) and for comparison the radial velocities from
cross-correlation for DRS extraction (\emph{bottom}).}
\end{figure}

We have chosen an asteroseismology run from the night 2004-10-02 for
the star Tau~Cet, which is a known RV standard with a dispersion
at the 1\,m/s level over years \citep{Pepe2011}. This night provides
the large number (438) of spectra needed to visualise the close performance
of FOX and DRS (Fig.~\ref{fig:RV-TauCet}). Since all data were taken
during the same night, we can defer discussions about the systematics
due to the wavelength calibration, which is another crucial step in
data reduction.

For the RV measurements we use the method of least-square template
matching \citep{Anglada2012}. The RVs are measured over several orders,
the ten bluest orders, as well as regions contaminated by telluric
absorptions features have been excluded. The top panel in Fig.~\ref{fig:RV-TauCet}
shows that RV differences resulting from the FOX and DRS extraction
are small and at or below the 1\,m/s level. The rms is 1.50\,m/s
with FOX and 1.49\,m/s with DRS extraction. Since the DRS pipeline
also delivers high-quality RVs measured by cross-correlation with
a binary template, we provide a comparison in the lower panel of Fig.~\ref{fig:RV-TauCet}
showing the impact of using the same extraction method (DRS) but a
different RV computation.

The two examples presented in this section demonstrate that the concept
of FOX can work in practice and indicate that FOX has a similar and
comparable performance in terms of S/N and RV precision compared to
standard optimal extraction.

\section{Limits of optimal extraction}

FOX and other optimal extraction algorithms assume 1D slit functions
that are aligned with the detector columns. However, in practice the
slit or fibre image is usually resolved and sampled by a few pixels
in the spatial and dispersion direction, and the injection of a monochromatic
wavelength causes a 2D PSF that can be seen in emission line spectra.
We investigate this effect in data from HARPS.

\begin{figure}
\centering

\includegraphics[bb=44bp 317bp 582bp 474bp,clip,width=1\linewidth]{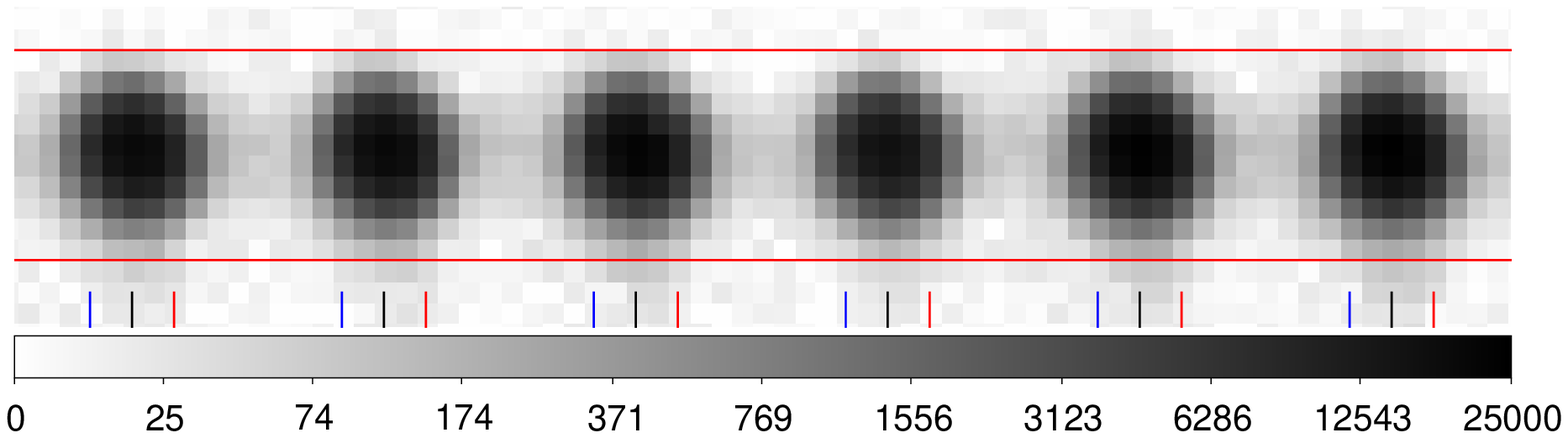}

\includegraphics[clip,width=1\linewidth]{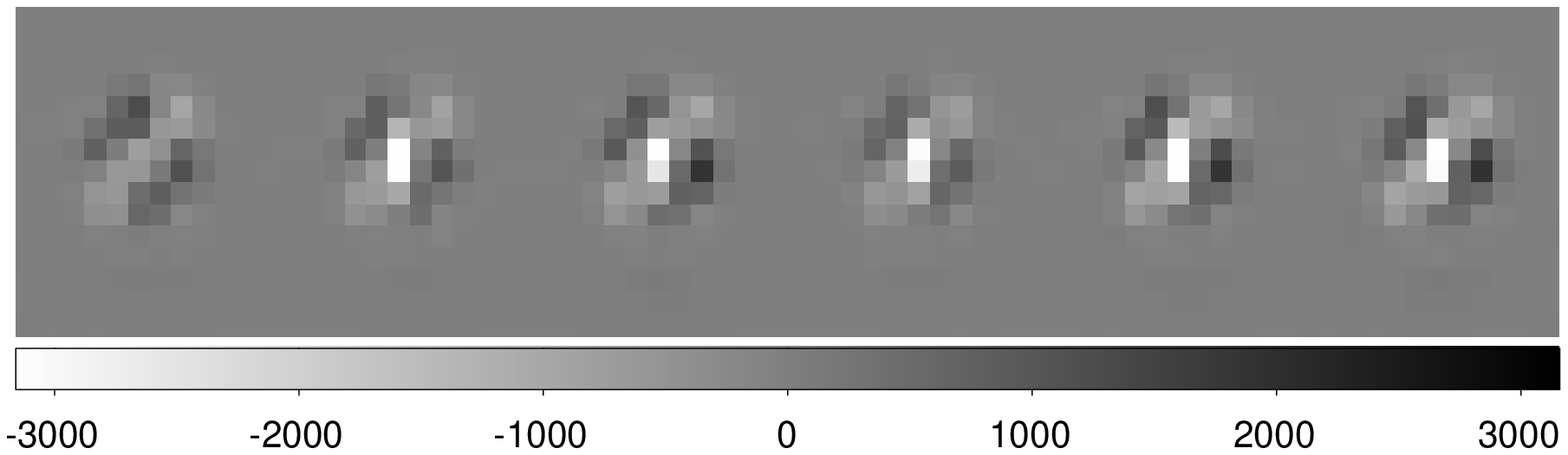}

\caption{\label{fig:LFC-residuals}Section of a HARPS laser frequency spectrum
$S_{x,y}$ (\emph{top}, logarithmic intensity scale) and residuals
$S_{x,y}-\hat{S}_{x,y}$ after the FOX extraction (\emph{bottom},
linear scale). Coloured ticks indicate positions of cross-section
cuts (see Fig.~\ref{fig:LFC-crosssections}).}
\end{figure}

Figure~\ref{fig:LFC-residuals} shows the sharp features of a spectrum
from a laser frequency comb (LFC; \citealp{Murphy2007,Wilken2010})
observed with HARPS (order 44). The residuals of the extraction ($S_{x,y}-\hat{S}_{x,y}$)
are shown in the lower panel, where $\hat{S}_{x,y}$ is computed with
the extracted spectrum $r_{x}$ and Eq.~(\ref{eq:FOXmodel_used})).
Of course, the residuals are larger in regions with larger flux (more
photon noise). However, a systematic (not random) pattern remains:
the residuals are always too low in the peak centre (by $\sim$10\%!)
and too high up left, as well as down right from the peak centres
(this pattern is also similar to the simulation of \citet{Bolton2010};
see their Fig.~1). The reason for this pattern is the mismatch between
the individual 2D PSF cross-sections and flat-field cross-profiles
(which could be thought of as the 2D PSF cross-sections integrated
along the disperion axis). Only if the individual 2D PSF cross-sections
were self-similar would the mismatch diminish.

\begin{figure}
\centering

\includegraphics[bb=50bp 50bp 673bp 390bp,clip,width=1\linewidth]{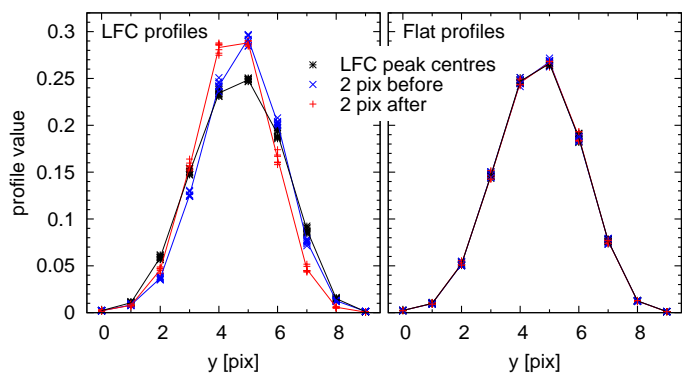}

\caption{\label{fig:LFC-crosssections}Cross-sections at different positions
in the laser frequency comb spectrum and flat-field image. The cut
positions are at the LFC peak maximum (black asterisks), two pixels
before (blue crosses), and two pixels after (red plus), as indicated
by the corresponding colored ticks in Fig.~\ref{fig:LFC-residuals};
the coloured lines connect the corresponding profile means. The profiles
are normalised to unit area.}
\end{figure}

We visualise some cross-sections in Fig.~\ref{fig:LFC-crosssections}
at the positions indicated in Fig.~\ref{fig:LFC-residuals}. This
region was selected because the order tilt is small and the LFC peak
distances are close to integer values (multiple of 12.0 pixels), i.e.
the LFC peaks have only small pixel phase shifts here and samples
similar phases of the effective PSF. Each of the 18 profiles (3 cuts
for each of the 6 peaks) is normalised to unit area. Obviously the
profiles in the flanks have a different form (more peaky and shifted)
with respect to the peak centres. The mismatches can be at the 10\%
level for sharp features (while they are not visible in the flat-field
cross-sections taken at the same positions). This will lead to high
$\chi_{\mathrm{red}}$-values in optimal extraction and biased extracted
spectrum values in sharp, high S/N features.

The visible, systematic residual tilts in Fig.~\ref{fig:LFC-residuals}
from bottom left to top right results from an asymmetry in the PSF,
although the PSF appears like a symmetric 2D Gaussian at first glance.
We presume that the physical explanation for this asymmetry (present
everywhere on the detector) lies in the quasi Littrow mode configuration,
i.e. the usage of R4 echelle grating with an off-plane angle of 1.5$^{\circ}$
that amplifies by a factor of $\approx8$ to a spectral line tilt
or shear \citep{Hearnshaw2009} and deforms the circularly symmetric
image of the fibre exit.

The shortcoming of ``optimal'' extraction might be solved with ``perfect''
extraction that involves a 2D PSF as outlined in \citet{Bolton2010}.
It can theoretically also deal with stray light and ghost features.
However, besides the increased computational effort, the main problem
in practice is to obtain the calibration matrix $\Psi_{x,y,\lambda}$.

\section{Conclusion}

We have introduced a method extracting 1D spectra from 2D raw data
using flat-field exposures as a measure of the instrumental profile.
This method is similar to standard optimum extraction. It does not
make any assumptions about the instrumental profile but requires its
temporal stability between flat-field and science exposures. The method
is well suited to stabilised fibre-fed spectrographs optimised for
high-precision radial velocity work.

One of the main advantages of FOX is that the reconstruction of the
instrumental PSF becomes unnecessary. Modelling of the PSF is a time-consuming
step, and in regions of low signal, the PSF is often ill-defined in
science exposures. It is therefore a strong advantage to take the
PSF from well-defined flat-field exposures, which is the main idea
of FOX. Following this scheme, extraction, masking, flat-fielding,
and blaze correction are all carried out in one step without any need
for data fitting when the calibration matrix is constructed. The decrease
in required CPU time is significant, which is particularly relevant
for large surveys like HARPS and CARMENES \citep{Quirrenbach2012}.
Furthermore, FOX has no requirements concerning the spectral format
(such as slowly varying spatial profiles), relaxes the need for accurate
localisation of spectral orders, and does not involve any numerical
unstable operations like division by the flat-field.

We compared the performance of FOX to standard optimal extraction
using HARPS data and the HARPS data reduction system. The results
are very similar with insignificant differences in the S/N. The spectra
of individual orders extracted with FOX match well in the overlap
regions showing that the inherent blaze correction works. We computed
the RV series from HARPS spectra extracted with FOX and DRS and find
them indistinguishable in terms of their rms scatter. We conclude
that FOX is a highly efficient and very robust method for extracting
astronomical spectroscopic data observed with stabilised fibre-fed
spectrographs. FOX cannot overcome the limitations caused by tilted
PSFs, stray light, or ghost features, but it can significantly improve
the robustness and time efficiency of existing and future data reduction
procedures.
\begin{acknowledgements}
MZ acknowledges support by the European Research Council under the
FP7 Starting Grant agreement number 279347. We thank the referee,
N. Piskunov, for many useful comments. We are also thankful to C.
Lovis and G. Lo Curto for helpful discussions about the HARPS instrument
and F. Bauer for reading the manuscript.
\end{acknowledgements}
\bibliographystyle{aa}
\bibliography{FOX}

\begin{appendix}

\section{\label{sec:Appendix}Appendix}

To estimate for each pixel the residual variance, 
\begin{equation}
\mathrm{var}(S_{x,y}-\hat{S}_{x,y})=\mathrm{var}(S_{x,y})+\mathrm{var}(\hat{S}_{x,y})-2\mathrm{cov}(S_{x,y},\hat{S}_{x,y}),\label{eq:res_var_def}
\end{equation}
we need the pixel variance, which is $\mathrm{var}(S_{x,y})=\sigma_{x,y}^{2}=\frac{1}{w_{x.y}}$,
and an expression for the model $\hat{S}_{x,y}$. Inserting Eqs.~(\ref{eq:FOX})
and (\ref{eq:error_abs}) into Eq.~(\ref{eq:FOXmodel_used}) gives
\begin{equation}
\hat{S}_{x,y}=F_{x,y}r_{x}=F_{x,y}\epsilon_{r_{x}}^{2}\sum_{y}w_{x,y}F_{x,y}S_{x,y}\,.\label{eq:Sxy-estimate}
\end{equation}
We see that $S_{x,y}$ itself contributes to $\hat{S}_{x,y}$ and
a covariance term will persist. Assuming that the pixels are otherwise
independent, the variance of the residuals is given by
\begin{align}
\mathrm{var}(S_{x,y}-\hat{S}_{x,y}) & =\sigma_{x,y}^{2}+F_{x,y}^{2}\mathrm{var}(r_{x})-2F_{x,y}^{2}\epsilon_{r_{x}}^{2}w_{x,y}\sigma_{x,y}^{2}\nonumber \\
 & =\sigma_{x,y}^{2}-F_{x,y}^{2}\epsilon_{r_{x}}^{2}\,.\label{eq:res_var}
\end{align}
This equation can be simplified for three special cases.

\emph{Case~1}. For pixels with relatively low profile weights ($F_{_{x,y}}^{2}\epsilon_{r_{x}}^{2}\ll\sigma_{x,y}^{2}$),
e.g. in the wings of spatial profiles, the residual variance just
becomes $\sigma_{x,y}^{2}$.

\emph{Case~2.} When the pixel variance is the same for all pixels
$\sigma_{x,y}=\sigma_{0}$, e.g. readout noise dominates, we have
($\epsilon_{r_{x}}^{2}=\sigma_{0}^{2}\frac{1}{\sum F_{x,y}^{2}}$)
\begin{align}
\left.\mathrm{var}(S_{x,y}-\hat{S}_{x,y})\right|_{\sigma_{x,y}=\sigma_{0}} & =\left[1-\frac{F_{x,y}^{2}}{\sum F_{x,y}^{2}}\right]\sigma^{2}\,.\label{eq:res_var_readout}
\end{align}
Moreover, for a uniform profile ($F_{x,y}=1$) the factor in the bracket
becomes $\frac{N-1}{N}$ (a well known correction factor for the unbiased
variance).

\emph{Case~3}. Assuming $\sigma_{x,y}^{2}=g\hat{S}_{x,y}=gr_{x}F_{x,y}$,
i.e. photon noise dominates, we find ($\epsilon_{r_{x}}^{2}=\frac{1}{\sum_{y}F_{x,y}}gr_{x}=\frac{1}{\sum_{y}F_{x,y}}\frac{\sigma_{x,y}^{2}}{F_{x,y}}$)
\begin{align}
\left.\mathrm{var}(S_{x,y}-\hat{S}_{x,y})\right|_{\sigma_{x,y}^{2}=g\hat{S}_{x,y}} & =\left[1-\frac{F_{x,y}}{\sum F_{x,y}}\right]\sigma_{x,y}^{2}\,.\label{eq:res_var_photon}
\end{align}
Again for a uniform profile the prefactor becomes $\frac{N-1}{N}$.

\end{appendix}
\end{document}